\date{\today}

\documentclass[conference]{IEEEtran}
\usepackage{balance}
\usepackage{standard}
\usepackage{etex}
\usepackage{rotate}
\usepackage{algorithmic}
\usepackage{algorithm}
\usepackage{float}
\usepackage{tikz}
\usepackage{refcount}
\usepackage{caption}
\usepackage{pgf,tikz}
\usepackage{pgfplots}
\usepackage{pgfplotstable}
\usepackage{bbm}
\usepackage{adjustbox}
\usetikzlibrary{calc}
\usepackage{relsize}
\usepackage{ifthen}
\usetikzlibrary{calc,fit,arrows,plotmarks,intersections,patterns,shapes,decorations,positioning, backgrounds, matrix}
\usepackage[numbers,sort&compress]{natbib}
\usepackage[ngerman, english]{babel}
\usetikzlibrary{arrows.meta}
\edef\wdArrowLength{2}
\tikzset{>={Latex[width=1.5mm,length=\wdArrowLength mm]}}
\usepackage{graphicx}
\usepackage[outdir=./pictures]{epstopdf}
\usepackage{cancel}
\usepackage{mathtools}
\usepackage{shadethm}
\usepackage[ansinew]{inputenc} % CHANGE HERE IF NECESSARY
\usepackage[T1]{fontenc}
\usepackage{empheq}
\usepackage{skull}
\usepgfplotslibrary{units}
\usepackage{calc}
\usepackage{nicefrac}
\usepackage{fancybox}
\usepackage{psfrag}
\usepackage{dsfont}
\newtheorem{prop}{Proposition}
\usepackage{pdfpages}
\usepackage{scrlayer-scrpage}
\usepackage{hyphenat}
\title{On Synergistic Benefits of Rate Splitting in IRS-assisted Cloud Radio Access Networks} % Title of the Thesis
\author{Kevin Weinberger, Alaa Alameer Ahmad, Aydin Sezgin\\ Ruhr-Universit\"at Bochum, Germany\\ Email:\{kevin.weinberger, alaa.alameerahmad, aydin.sezgin\}@rub.de,}
\date{\today}

\usepackage{graphicx}
\usepackage{placeins}
\usepackage{booktabs}
\usepackage{marvosym}
\usepackage{multirow}

\usepackage{latexsym, amsmath, amssymb, amsfonts, upgreek}
\usepackage[nolist]{acronym}

\usepackage[babel,english=american, german=quotes]{csquotes}

\usepackage{tikz}
\usetikzlibrary{calc,arrows,positioning,decorations,shadows,shapes,fadings,matrix}
\tikzset{>=latex'}
\tikzset{semithick}

\makeatletter
\providecommand{\IfElsePackageLoaded}[3]{\@ifpackageloaded{#1}{#2}{#3}}
\makeatother

\usepackage{subfigure}
\IfElsePackageLoaded{subfig}
	{\usepackage[subfigure]{tocloft}}{	
	\IfElsePackageLoaded{subfigure}
		{\usepackage[subfigure]{tocloft}}
		{\usepackage{tocloft}}
	}

\definecolor{colKeys}{rgb}{0,0,1}
\definecolor{colIdentifier}{rgb}{0,0,0}
\definecolor{colComments}{rgb}{1,0,0}
\definecolor{colString}{rgb}{0,0.5,0}

\usepackage{listings}
\tikzstyle{dot} = [draw, circle, minimum size=0.2pt, scale=0.3, fill=black]
\tikzstyle{state} = [draw, circle, minimum width=1cm]
\tikzstyle{myloop} = [->,to path={.. controls +(45:1) and +(135:1) .. (\tikztotarget) \tikztonodes}]
\tikzstyle{block} = [draw, fill=blue!20, rectangle, rounded corners, drop shadow, minimum height=3em, minimum width=6em]
\tikzstyle{pc-or} = [draw, fill=blue!20, rectangle, rounded corners, drop shadow, minimum height=6em, minimum width=5em]
\tikzstyle{scan-or} = [draw, fill=red!80, rectangle, rounded corners, drop shadow, minimum height=6em, minimum width=5em]
\tikzstyle{dot2} = [draw, circle, minimum size=0.2pt,scale=0.3,fill=black,black]
\tikzstyle{ant} = [regular polygon,regular polygon sides=3, draw,shape border rotate=180,minimum size=0.2pt,scale=0.3]
\tikzstyle{connector} = [->,thick]
\tikzstyle{snakeline} = [connector, decorate, decoration={pre length=0.2cm, post length=0.2cm, snake, amplitude=.4mm, segment length=2mm}]

\usepackage{pgfplots}
\usepackage{pgfplotstable}
\usetikzlibrary{spy}

\usepackage{scrlayer-scrpage}
\ofoot{\pagemark}
\ohead{\headmark}
\usepackage{shadethm}
\newshadetheorem{sthm}{Definition}[chapter]

\usepackage[breaklinks,hyperindex,colorlinks,anchorcolor=black,citecolor=black,filecolor=black,linkcolor=black,menucolor=black,urlcolor=black,pdftex]{hyperref} % pagebackref: Add page number to the references where they can be found
\usepackage{breakurl}

\makeatletter
\AtBeginDocument{
 \hypersetup{
   pdftitle = {\@title},
   pdfauthor = {\@author},
   pdfsubject={\@title},
   pdfkeywords={Generating Secure Cryptographic Keys Channel Properties Matlab Simulink}, % CHANGE HERE
 }
}

\def\tikz@delimiter#1#2#3#4#5#6#7#8{%
	\bgroup
		\pgfextra{\let\tikz@save@last@fig@name=\tikz@last@fig@name}%
		node[outer sep=0pt,inner sep=0pt,draw=none,fill=none,anchor=#1,at=(\tikz@last@fig@name.#2),#3]
		{%
			{\nullfont\pgf@process{\pgfpointdiff{\pgfpointanchor{\tikz@last@fig@name}{#4}}{\pgfpointanchor{\tikz@last@fig@name}{#5}}}}%
			\delimitershortfall\z@% as suggested by morbusg (maximum space not covered by a delimiter = 0)
			\resizebox*{!}{#8}{$\left#6\vcenter{\hrule height .5#8 depth .5#8 width0pt}\right#7$}%
		}
		\pgfextra{\global\let\tikz@last@fig@name=\tikz@save@last@fig@name}%
	\egroup%
}

\tikzset{hexagon/.code={
	\draw (0,2) -- (-4,0) -- (0,-2) -- (4,0) -- (0,2);
}}

\tikzset{phone/.code={
   \node [rectangle,rounded corners=1.5pt,draw,minimum height=0.6cm, minimum width=0.35cm] at (0,0){};
   \node [rectangle,rounded corners=1.5pt,draw,minimum height=0.5cm, minimum width=0.3cm] at (0,0){};
}}

\makeatletter
\def\cantox@vector#1#2#3#4#5#6#7#8{%
  \dimen@.5\p@
  \setbox\z@\vbox{\boxmaxdepth.5\p@
   \hbox{\kern-1.2\p@\kern#1\dimen@$#7{#8}\m@th$}}%
  \ifx\canto@fil\hidewidth  \wd\z@\z@ \else \kern-#6\unitlength \fi
  \ooalign{%
    \canto@fil$\m@th \CancelColor
    \vcenter{\hbox{\dimen@#6\unitlength \kern\dimen@
      \multiply\dimen@#4\divide\dimen@#3 \vrule\@depth\dimen@\@width\z@
      \vector(#3,-#4){#5}%
    }}_{\raise-#2\dimen@\copy\z@\kern-\scriptspace}$%
    \canto@fil \cr
    \hfil \box\@tempboxa \kern\wd\z@ \hfil \cr}}
\def\bcancelto#1#2{\let\canto@vector\cantox@vector\cancelto{#1}{#2}}
\makeatother

\newcommand{\ifthen}[2]{\ifthenelse{#1}{#2}{}}

\newcommand{\myNorm}[1]{\left\lVert#1\right\rVert}

\newcommand{\sqr}[1]{{#1}^2}
\newcommand{\sbrackets}[1]{\left(#1\right)}
\newcommand{\hbrackets}[1]{\left[#1\right]}
\newcommand{\vectw}{\vect{\omega}}
\newcommand{\vectv}{\vect{v}}
\newcommand{\vectvt}{\tilde{\vect{v}}}
\newcommand{\matV}{\mat{V}}

\newcommand{\vecth}{\vect{h}}
\newcommand{\matH}{\mat{H}}
\newcommand{\vecta}{\vect{a}}
\newcommand{\matM}{\mat{M}}

\newcommand{\cbrackets}[1]{\left\{#1\right\}}
\newcommand{\Tr}[1]{\text{Tr}\left(#1\right)}
\renewcommand{\rank}[1]{\text{rank}\left(#1\right)}

\newcommand{\listequationsname}{List of Formulas}
\newlistof{myequations}{equ}{\listequationsname}

\definecolor{myblue1}{rgb}{0,0,255}
\definecolor{myblue2}{rgb}{65,105,225}
\definecolor{myblue3}{rgb}{70,130,180}
\definecolor{myblue4}{rgb}{176,196,222}

\newcommand{\mytilde}{{\raise.17ex\hbox{$\scriptstyle\mathtt{\sim}$}}}
\newcommand{\naive}{}
\def\naive/{na\"{\i}ve}

\newcommand{\executeiffilenewer}[3]{%
\ifnum\pdfstrcmp{\pdffilemoddate{#1}}%
{\pdffilemoddate{#2}}>0%
{\immediate\write18{#3}}\fi%
}
\pgfkeys{/pgf/fpu}
\pgfmathparse{16383+1}

\pgfkeys{/pgf/fpu=false}

\def\endthebibliography{%
  \def\@noitemerr{\@latex@warning{Empty `thebibliography' environment}}%
  \endlist
}
\begin{document}
\maketitle
\begin{abstract}
The concept of \acp{IRS} is considered as a promising technology for increasing the efficiency of mobile wireless networks. This is achieved by employing a vast amount of low-cost individually adjustable passive reflect elements, that are able to apply changes to the reflected signal. To this end, the \ac{IRS} makes the environment real-time controllable and can be adjusted to significantly increase the received signal quality at the users by passive beamsteering. However, the changes to the reflected signals have an effect on all users near the \ac{IRS}, which makes it impossible to optimize the changes to positively influence every transmission, affected by the reflections. This results in some users not only experiencing better signal quality, but also an increase in received interference. To mitigate this negative side effect of the \ac{IRS}, this paper utilizes the \ac{RS} technique, which enables the mitigation of interference within the network in such a way that it also mitigates the increased interference caused by the \ac{IRS}. To investigate the effects on the overall power savings, that can be achieved by combining both techniques, we minimize the required transmit power, needed to satisfy per-user \ac{QoS} constraints. Numerical results show the improved power savings, that can be gained by utilizing the \ac{IRS} and the \ac{RS} technique simultaneously. In fact, the concurrent use of both techniques yields power savings, which are beyond the cumulative power savings of using each technique separately.
\end{abstract} 
\thispagestyle{empty}
\pagestyle{empty}
\section{Introduction}
With the introduction of solutions based on the \ac{IoT} on various application areas, such as healthcare, industrial control etc., billions of new devices with different applications will connect to \ac{B5G} wireless networks \cite{microsoftIoT}, which naturally results in an increase in interference. Furthermore, more \acp{BS} are deployed within each cell, due to the utilization of the spatial reuse technique, which also increases inter-cell interference. To tackle the challenges of modern communication networks, \acp{IRS} have been shown to be a promising cost-effective solution to increase the achievable data rates of existing wireless networks by improving the spectral efficiency and the signal coverage \citep{,IRS_intr2,IRS_intr4}. The \ac{IRS} is a physical reflective metasurface consisting of small reflect elements, whose changes to the reflected signal can be controlled independently by a smart controller. The deployment of an \ac{IRS} therefore enables the possibility to smartly reflect incident waves and simultaneously adjust the phase shift of each reflect element. Consequently, the \ac{IRS} is able to add phase shifts at the user either constructively, thus improving the received signal power, or destructively, and thus mitigating interference. However, the changes to the reflected signals do not result in an improvement for all \ac{IRS}-assisted transmissions. Therefore, the \ac{IRS} is not only amplifying the received signal, but also the received interference for most users, especially in dense networks. 

For this reason, this paper utilizes the \ac{RS} technique, which is able to  mitigate interference within the network in such a way that the interference, amplified by the \ac{IRS}, is also mitigated. To enable the practical implementation of the \ac{RS} scheme, a \ac{C-RAN} setting is adopted \citep{Alaa_EE_RSCMD,Alaa_27}, in which the central processor at the cloud is connected to a set of \acp{BS} via high-speed and high-capacity fronthaul links \cite{C-RAN}, while simultaneously adjusting the \ac{IRS}. 

To study the effect of both techniques on the minimum required transmit power, we investigate the optimal power control in the \ac{IRS}-assisted and RS-enabled \ac{C-RAN} to minimize the required transmit power under per-user minimum \ac{QoS} constraints. This poses a challenging optimization problem due to the coupling of the optimization variables in the non-convex \ac{QoS} constraints. Therefore, an alternating optimization framework is proposed, which decouples the variables and solves the emerging sub-problems alternatively. The results obtained are compared with a baseline scheme of \ac{TIN} \citep{5074583,jabar}.
\section{System Model}\label{ch:Sysmod}

\begin{figure}
\centering
\includegraphics[width=\columnwidth]{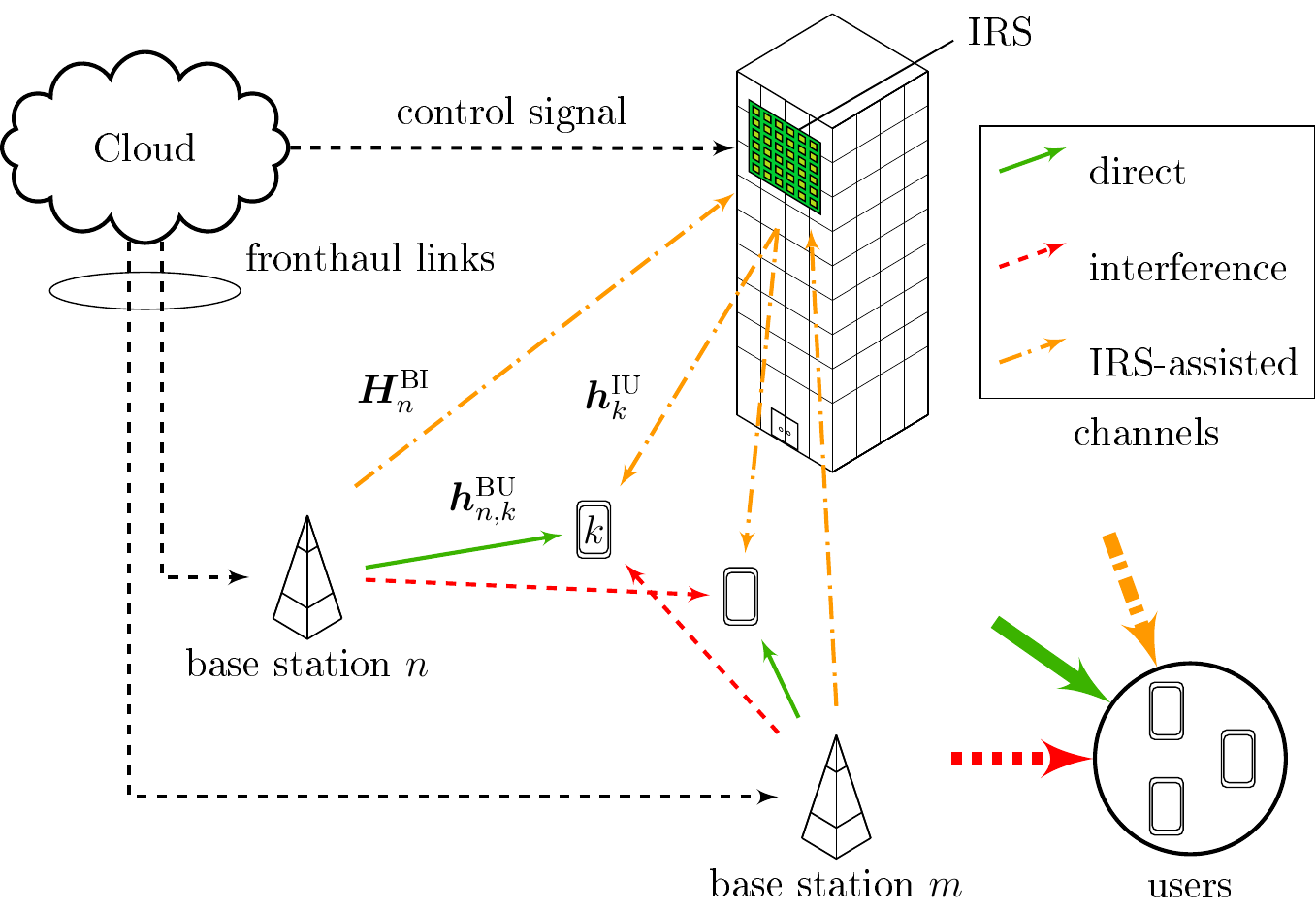}
\caption{IRS-assisted multiuser \ac{C-RAN} system}
\label{fig:IRS-CRAN}
\end{figure}

The system model considered in this work is depicted in Figure \ref{fig:IRS-CRAN} and consists of a \ac{RS}-enabled and \ac{IRS}-assisted \ac{C-RAN} downlink system. In more details, the network consists of a set of multi-antenna \acp{BS} $\mathcal{N} = \{1,2,\cdots,N\}$, each of which is equipped with $L \geq 1$ antennas. The \acp{BS} serve a set of single-antenna users $\mathcal{K} = \{1,2,\cdots,K\}$. An \ac{IRS}, composed of $R$ passive real-time-controllable reflect elements, is deployed in the communication environment to assist the \acp{BS} in their communication with the users. Each \ac{BS} $n \in \mathcal{N}$ is connected to the \ac{CP} at the cloud via orthogonal fronthaul links of unlimited capacity. Each user $k$ requires to be served with a minimum data rate $r_k^{\text{Min}}$, which represents the \ac{QoS} target of user $k$.

The channel vector, denoted by ${\vect{h}_{n,k} \in \mathbb{C}^{L\times1}}$, represents the direct channel links between \ac{BS} $n$ and user $k$. The channels of the \ac{IRS}-assisted path are composed of $\mat{H}_{n}^\text{BI} \in \mathbb{C}^{L \times R}$ and $\vecth_{k}^\text{IU} \in \mathbb{C}^{R \times 1}$ as depicted in Figure \ref{fig:IRS-CRAN}.
Let $\vect{h}_k = \left[ ({\vect{h}_{1,k}})^{\,T } ,({\vect{h}_{2,k}})^T, \dots ,({\vect{h}_{N,k}})^T \right]^T  \in \mathbb{C}^{NL\times1}$ be the aggregate direct channel vector of user $k$, $\vect{H}^{\text{BI}} = \left[ ({\vect{H}_{1}^{\text{BI}}})^{\,T } ,({\vect{H}_{2}^{\text{BI}}})^T, \dots ,({\vect{H}_{N}^{\text{BI}}})^T \right]^T  \in \mathbb{C}^{NL\times R}$ be the aggregate channel matrix from the \acp{BS} to the \ac{IRS} and $\vect{x} = \left[ {\vect{x}_{1}}^{\,T } ,{\vect{x}_{2}}^T, \dots , {\vect{x}_{N}}^T \right]^T  \in \mathbb{C}^{NL\times1}$ be the aggregate transmit signal vector. Using this notation, the received signal at user $k$ can be written as
\begin{align}\label{eq:recSig}
    y_k & = (\vect{h}_{k})^H\, \vect{x} + (\mat{H}^\text{BI} \, \mat{\Theta} \, \vect{h}_{k}^{\text{IU}})^H \, \vect{x} + n_k,
%       &= (\vect{h}_{k}^\text{BU} + \mat{H}^\text{BI} \, \mat{\Theta} \, \vect{h}_{k}^{\text{IU}})^H \, \vect{x} + n_k, \label{eq:recSig_pre}\\
\end{align}
where $n_k \sim \mathcal{CN}(0,\sigma^2)$ is the \ac{AWGN}, $\mat{\Theta}\, = \text{diag}(\vectv) \in \mathbb{C}^{R \times R}$ is a diagonal reflection coefficient matrix accounting for the response of the reflect elements. Let the phase shift vector be defined as $\vectv = [v_{1}, v_{2} , \dots v_{R}]^T$, where each reflection coefficient $v_r$ consists of a phase shift $\theta \in [0,2\pi]$, namely $v_r = e^{j\theta_r}$. By introducing the matrix $\matH_k = \matH^{\text{BI}}\, \text{diag}(\vecth_k^{\text{IU}})$ and denoting the sum of the aggregated direct and reflected channel vectors of user $k$ as an effective channel
$\vecth_k^{\text{eff}}(\vectv) = \vect{h}_{k} + \mat{H}_k \vectv$, the received signal (\ref{eq:recSig}) can be rewritten as $y_k  = (\vect{h}_k^{\text{eff}}(\vectv))^H \, \vect{x} + n_k$,

\subsection{Rate Splitting}
The \ac{CP} splits $q_k$, the requested message of user $k$, into two sub-messages, namely a private part $q_k^p$ and a common part $q_k^c$. Subsequently, the respective parts are encoded by the \ac{CP} into the private and common symbols $s_k^p$ and $s_k^c$, respectively. The coded symbols $s_k^p$ and $s_k^c$ are assumed to form an \ac{i.i.d.} Gaussian codebook. The \ac{CP} shares the private symbols $s_k^p$ (common symbols $s_k^c$) with a cluster of predetermined \acp{BS}, that exclusively send the beamformed private (common) symbols to user $k$. Given these definitions, the subset of users that are served by \ac{BS} $n$ with a private or common message $\mathcal{K}_n^p,\,\mathcal{K}_n^c \subseteq \mathcal{K}$, respectively, are defined by
\begin{align}
   \mathcal{K}_n^p &= \{k\in\mathcal{K} \, | \, \text{BS } n \text{ serves } s_k^p \text{ to user }k\},\\
   \mathcal{K}_n^c &= \{k\in\mathcal{K} \, | \, \text{BS } n \text{ serves } s_k^c \text{ to user }k\}.
\end{align}
The beamformers $\vectw_{n,k}^p$ and $\vectw_{n,k}^c$ used by \ac{BS} $n$ to send $s_k^p$ and $s_k^c$, respectively,  are created by the \ac{CP} and forwarded to \ac{BS} $n$ through the fronthaul link along with the respective private symbols $\cbrackets{s_k^p \,|\, \forall k \in \mathcal{K}_n^p}$ and common symbols $\cbrackets{s_k^c \,|\, \forall k \in \mathcal{K}_n^c}$.
After \ac{BS} $n$ receives the corresponding messages and beamformers, it constructs the transmit signal vector $\vect{x}_n$ and sends it to the users of interest. By denoting the aggregate beamforming vectors as $\vectw_k^o = \left[ (\vectw_{1,k}^o)^T,(\vectw_{2,k}^o)^T,\dots,(\vectw_{N,k}^o)^T  \right]^T  \in \mathbb{C}^{NL\times1}, \forall \, o \in \{p,c\}$ associated with $s_k^p$ and $s_k^c$, respectively, the aggregate transmit signal vector can be expressed as \begin{align}
\vect{x} = \sum_{k \in \mathcal{K}_n^p}\vectw_k^p s_k^p + \sum_{k \in \mathcal{K}_n^c}\vectw_k^c s_k^c.
\end{align}
Consequently, should \ac{BS} $n$ not participate in the cooperative transmission of the private or common symbol of user $k$, the respective beamformers are set to zero, namely $\vectw_{n,k}^p = \zero_L$ or $\vectw_{n,k}^c = \zero_L$, where $\zero_L$ denotes a column vector of length $L$ with all zero entries.

\subsection{Achievable rates}
In this work, the influence of the \ac{CMD} scheme, adopted by the users, is utilized for the purpose of interference mitigation. Hence, a successive decoding strategy is adopted, in which user $k$ decodes the common message of the user with the strongest channel (in the Euclidean norm sense) first. Let $\mathcal{M}_k$ be the set of users that decode $s_k^c$, i.e.,
\begin{align}
\mathcal{M}_k = \cbrackets{j \in \mathcal{K} \,|\, \text{ user } j \text{ decodes }s_k^c}.
\end{align}
In conjunction, let the set of users $\Phi_k$, whose common messages are decoded by user $k$, and the set of users $\Psi_k$, whose common messages are \textit{not} decoded by user $k$, be defined as
\begin{align}
\Phi_k = \cbrackets{j \in \mathcal{K} \, | \, k \in\mathcal{M}_j},\, \Psi_k = \cbrackets{j \notin \mathcal{K} \, | \, k \in\mathcal{M}_j},
\end{align}
respectively. It is noted that $\Phi_k$ and $\Psi_k$ are two disjoint subsets from the set of active users $\mathcal{K}$, while the cardinality of $\Phi_k$ is bounded by $D$, i.e., $|\Phi_k| \leq D.$ For the sets $\Phi_k ,\, \forall k\in\mathcal{K}$ a decoding order is established, which is represented by a mapping of an ordered set with cardinality of $|\Phi_k|$, i.e.,
\begin{align}
   \pi_k(j) : \cbrackets{1,2,\dots,|\Phi_k|} \rightarrow \Phi_k.
\end{align}
Assuming user $k$ decodes the common messages of user $j_1$ and user $j_2$ with $j_1 \neq j_2$, then $\pi_k(j_1) > \pi_k(j_2)$ implies that user $k$ decodes the common message of user $j_1$ first and the common message of user $j_2$ afterwards.
Furthermore, the set $\Omega_{i,k}$ is defined as
\begin{align}
   \Omega_{i,k} = \cbrackets{m\in\Phi_i \, | \, \pi_i(k) > \pi_i(m)}
\end{align}
and represents the set of users whose common messages are decoded by user $i$ after decoding the common message of user $k$.

Now, the received signal at user $k$ can be expressed as
\begin{align}\label{eq:recSig_final}
  &y_k = \overbrace{\sbrackets{\vecth_k^{\text{eff}}(\vectv)}^H \vectw_k^p s_k^p + \sum_{j \in \Phi_k} \sbrackets{\vecth_k^{\text{eff}}(\vectv)}^H \vectw_j^c s_j^c}^{\text{signals that are decoded}}+ \nonumber\\ &   \underbrace{\sum_{m \in \mathcal{K}\text{\textbackslash}\cbrackets{k}} \hspace{-0.35cm} \sbrackets{\vecth_k^{\text{eff}}(\vectv)}^H \vectw_m^p s_m^p \hspace{-0.05cm} +\hspace{-0.05cm} \sum_{\ell \in \Psi_k} \sbrackets{\vecth_k^{\text{eff}}(\vectv)}^H \vectw_\ell^c s_\ell^c + n_k}_{\text{interference plus noise}}.
\end{align}
Let ${\gamma_k^p}$ denote the \ac{SINR} of user $k$, decoding its private message, and let $\gamma_{i,k}^c$ denote the \ac{SINR} of user $i$, decoding the common message of user $k$. Let the vector $\vectw_k = \hbrackets{ \sbrackets{\vectw_k^p}^T , \sbrackets{\vectw_k^c}^T }^T \in \mathbb{C}^{2NL\times1}$ represent the stacked private and common beamformers of user $k$ and let the vector $\vectw = \hbrackets{ (\vectw_1)^T , (\vectw_2)^T , \dots , (\vectw_K)^T}^T \in \mathbb{C}^{2KNL\times1}$ represent all beamformers.\\
Using the partition in (\ref{eq:recSig_final}), $\gamma_k^p(\vectw,\vectv)$ and $\gamma_{i,k}^c(\vectw,\vectv)$ can be expressed as
\begin{align}
   &\gamma_k^p(\vectw,\vectv) = \label{eq:gammap}\\
   &\frac{|{\sbrackets{\vecth_k^{\text{eff}}(\vectv)}^H \vectw_k^p}|^2}
                     {\sum\limits_{m \in \mathcal{K}\text{\textbackslash}\cbrackets{k}} \hspace{-0.2cm} |{\sbrackets{\vecth_k^{\text{eff}}(\vectv)}|^H \vectw_m^p}^2 + \sum\limits_{\ell \in \Psi_k} |{\sbrackets{\vecth_k^{\text{eff}}(\vectv)}|^H \vectw_\ell^c}^2 + \sigma^2}, \nonumber\\
   &\gamma_{i,k}^c(\vectw,\vectv) = \label{eq:gammac} \\[-1pt] &\frac{|{\sbrackets{\vecth_i^{\text{eff}}(\vectv)}^H \vectw_k^c}|^2}
                     {T_i + \sum\limits_{\ell \in \Psi_i} |{\sbrackets{\vecth_i^{\text{eff}}(\vectv)}|^H \vectw_\ell^c}^2 +
                     \sum\limits_{m \in \Omega_{i,k}} |{\sbrackets{\vecth_i^{\text{eff}}(\vectv)}|^H \vectw_m^c}^2},\nonumber
   \vspace{-0.1cm}
\end{align}
where $T_i = \sum_{j \in \mathcal{K}}^{\phantom{I}} |\sbrackets{\vecth_i^{\text{eff}}(\vectv)}^H \vect{\omega}_j^p|^2  + \sigma^2$.
Furthermore, the sum-rate of the common and private rates can be defined, which has to be higher or equal than the \ac{QoS} $r_k^{\mathsf{Min}}$ of each user $k$, namely \vspace{-0.1cm}
\begin{align}
   R_k^p + R_k^c \geq r_k^{\mathsf{Min}},
\end{align}
where $R_k^p$ is the private rate and $R_k^c$ the common rate of user $k$.
The messages $s_k^p$ and $s_k^c$ are only decoded reliably if the private and common rates of each user satisfy the following conditions
\begin{align}\label{eq:privateRate}
 R_k^p &\leq   B \log_2 \big(1+\gamma_{k\phantom{.}}^p(\vectw,\vectv)\big)  , \quad \forall k \in \mathcal{K}\\
  R_k^c &\leq B \log_2 \sbrackets{1+\gamma_{i,k}^c(\vectw,\vectv) }, \quad  \forall k \in \mathcal{K} ,\, \forall i \in \mathcal{M}_k,\label{eq:commonRate}
\end{align}
where $B$ represents the transmission bandwidth.

\section{Problem Formulation}\label{ch:propform}
Let the total transmit power be defined as
\begin{align}\label{eq:Ptr}
  P(\vectw) & =  \sum_{k\in \mathcal{K}}  \alpha_k \left( \sum_{n\in \mathcal{N}} \myNorm{ \vect{\omega}_{n,k}^p}_2^2 \ + \myNorm{ \vect{\omega}_{n,k}^c}_2^2 \right),
\end{align}
where $\alpha_k$ is a weighting factor, which represents the priority of serving user $k$ compared to other users in the network.
\noindent
The problem can then be defined as
\begin{align}\label{eq:genProb}
   &\underset{\vectw,\vectv}{\text{minimize}} \quad P(\vectw) \tag{P1}\\
   &\text{subject to} \quad \nonumber \\ &\log_2\sbrackets{1+\gamma_k^p(\vectw,\vectv)}+ \nonumber\\
   &\underset{i \in \mathcal{M}_k}{\min} \cbrackets{ \log_2\sbrackets{1+\gamma_{i,k}^c\sbrackets{\vectw,\vectv}}} \geq r_k^{\text{Min}}/B ,\,   \forall k \in \mathcal{K}\label{eq:general_Minrate}\\
   &|v_r| = 1, \hspace{3.75cm}  \forall r \in \cbrackets{1,...,R},\label{eq:general_IRS}
\end{align}
where the constraints $0\leq \theta_r \leq 2\pi, \forall r\in\{1,...,R\}$ are represented by the equivalent unit modulus constraints in (\ref{eq:general_IRS}).
The problem is generally difficult to solve, due to the non-convex constraints (\ref{eq:general_Minrate}) and (\ref{eq:general_IRS}). Furthermore, the optimization variables are coupled in the \ac{QoS}$\underset{\phantom{|}}{ }$constraints (\ref{eq:general_Minrate}).

\section{Alternating Optimization}
To facilitate practical implementation, an alternating optimization framework is introduced to solve the problem in a distributed fashion. To this end, problem (\ref{eq:genProb}) is first reformulated into a form suitable for the inner-convex approximation framework \citep{AlaaRSCMD,Alaa_EE_RSCMD}, and second recasted into an semidefinite programming (SDP) problem.
\subsection{Beamforming Design}
In the initial step of the algorithm, $\vectv$ is assumed to be fixed, and consequently not an optimization variable. The optimization problem (\ref{eq:genProb}) can therefore be reformulated to
\begin{align}\label{eq:genW}
    &\underset{\{\vect{w}_k,\vect{t}_k,\mat{R}_k\}_{k=1}^K}{\text{minimize}} \qquad P(\vectw) \tag{P2.1}\\
    &\text{subject to} \quad  \nonumber \\
    &R_k^p + R_k^c \geq r_k^{Min} \label{eq:p2.1_first}\\
%    &\sum_{k \in \mathcal{K}_n^p}R_k^p + \sum_{k \in \mathcal{K}_n^c}R_k^c \leq C_n \,,\qquad \forall n \in \mathcal{N}\\
    & R_k^p - B \log_2(1+t_k^p) \leq 0\\
    & R_k^c - B \log_2(1+t_k^c) \leq 0\\
   &\vect{t}_k \geq 0\\
   &\mat{R}_k \geq 0 \label{eq:2.1_last}\\
   & t_k^p \leq  \gamma_k^p(\vectw),     \hspace{2.35cm} \forall k \in \mathcal{K} \label{eq:tkp}\\
   & t_k^c \leq  \gamma_{i,k}^c(\vectw), \hspace{2.175cm} \forall k \in \mathcal{K} ,\, \forall i \in \mathcal{M}_k, \label{eq:tkc}
\end{align}
where the variables $\mat{R}_k= \hbrackets{R_k^p,R_k^c}^T$ and $\vect{t}_k=\hbrackets{t_k^p,t_k^c}^T$ are introduced. $\vect{t}_k  \geq  0$ and $\mat{R}_k  \geq  0$ indicate that vector $\vect{t}_k$ and matrix $\mat{R}_k$ are greater than or equal to 0 in a component-wise manner.
The constraints (\ref{eq:tkp}) and (\ref{eq:tkc}) still define a non-convex feasible set, which makes problem (\ref{eq:genW}) non-convex. To overcome this challenge, equal representations of the constraints (\ref{eq:tkp}) and (\ref{eq:tkc}) are adopted, namely
\begin{align}\label{eq:tkp2}
         & {\sum_{m\in\mathcal{K}\text{\textbackslash}\cbrackets{k}}^{K} |\sbrackets{\vecth_k^{\text{eff}}}^H \vect{\omega}_m^p|^2 + \sum_{\ell \in \Psi_k} |\sbrackets{\vecth_k^{\text{eff}}}^H \vect{\omega}_\ell^c|^2 + \sigma^2}\, \nonumber \\ & \hspace{3.8cm}-\, {\frac{|\sbrackets{\vecth_k^{\text{eff}}}^H \vect{\omega}_k^p|^2}{t_k^p}}\leq 0
         \end{align}
         \begin{align}
         &{T_i + \sum_{\ell \in \Psi_i} |\sbrackets{\vecth_i^{\text{eff}}}^H \vect{\omega}_\ell^c|^2 \phantom{\frac{|}{|}} \hspace{-0.075cm} + \sum_{m \in \Omega_{i,k}} |\sbrackets{\vecth_i^{\text{eff}}}^H \vect{\omega}_m^c|^2}\,\ \nonumber \\ & \hspace{3.8cm} -\, {\frac{|\sbrackets{\vecth_i^{\text{eff}}}^H \vect{\omega}_k^c|^2}
         {t_k^c}} \leq 0 \label{eq:tkc2}    ,
   \end{align}
where with an abuse of notation the dependency of $\vect{h}_k^{\text{eff}}$ from $\vectv$ is skipped. The constraints are now represented in the form of difference-of-convex functions, which makes it possible to approximate them by using the first-order Taylor approximation around a feasible point $(\tilde{\vectw},\tilde{\vect{t}})$ \cite{convexFunct2,boyd_convex}.

To derive a lower bound for the second (negative) convex terms, the first-order Taylor approximation is applied.$\underset{\phantom{|}}{ }$ Therefore,
if $(\tilde{\vectw},\overset{\phantom{,}}{\tilde{\vect{t}})}$ is a feasible point of problem (\ref{eq:genW}) then it holds \cite{AlaaRSCMD}, that
\begin{align}\label{DC-approxp}
\frac{|\sbrackets{\vecth_k^{\text{eff}}}^H \vect{\omega}_k^p|^2}
{t_k^p} \geq&
\frac{2\Re\cbrackets{ \sbrackets{\tilde{\vect{\omega}}_k^p}^H \vecth_k^{\text{eff}} \sbrackets{\vecth_k^{\text{eff}}}^H \vect{\omega}_k^p}}
{\overset{}{\tilde{t}_k^p}} \nonumber \\
\hspace{3cm}&-
\frac{|\sbrackets{\vecth_k^{\text{eff}}}^H \overset{\phantom{I}}{\tilde{\vect{\omega}}_k^p}|^2}
{\sbrackets{\tilde{t}_k^p}^2} \, t_k^p
\end{align}
and
\begin{align}\label{DC-approxc}
\hspace{0.05cm}\frac{|\sbrackets{\vecth_i^{\text{eff}}}^H \vect{\omega}_k^c|^2}
{t_k^c} \geq&
\frac{2\Re\cbrackets{ \sbrackets{\tilde{\vect{\omega}}_k^c}^H \vecth_i^{\text{eff}} \sbrackets{\vecth_i^{\text{eff}}}^H \vect{\omega}_k^c}}
{\overset{}{\tilde{t}_k^c}} \nonumber \\
\hspace{3cm}&-
\frac{|\sbrackets{\vecth_i^{\text{eff}}}^H \overset{\phantom{I}}{\tilde{\vect{\omega}}_k^c}|^2}
{\sbrackets{\tilde{t}_k^c}^2} \, t_k^c ,
\end{align}
where $\text{Re}\cbrackets{\cdot}$ denotes the real part of a complex-valued number.
The approximations (\ref{DC-approxp}) and (\ref{DC-approxc}) can be utilized to establish inner-convex subsets by substituting the corresponding terms in (\ref{eq:tkp2}) and (\ref{eq:tkc2}) with their respective lower bound, namely
\begin{align}
      &0 \geq \sum_{j\in\mathcal{K}\text{\textbackslash}\cbrackets{k}} |\sbrackets{\vecth_k^{\text{eff}}}^H \vect{\omega}_j^p|^2 + \sum_{\ell \in \Psi_k} |\sbrackets{\vecth_k^{\text{eff}}}^H \vect{\omega}_\ell^c|^2 + \sigma^2-
      \nonumber\\
      &\frac{2\Re\cbrackets{ \sbrackets{\tilde{\vect{\omega}}_k^p}^H \vecth_k^{\text{eff}} \sbrackets{\vecth_k^{\text{eff}}}^H \vect{\omega}_k^p}}
      {\overset{}{\tilde{t}_k^p}}+\frac{|\sbrackets{\vecth_k^{\text{eff}}}^H \overset{\phantom{I}}{\tilde{\vect{\omega}}_k^p}|^2}
      {\sbrackets{\tilde{t}_k^p}^2} \, t_k^p,\forall k \in \mathcal{K}  \nonumber \\
       \label{eq:approxCon1}\\
      &0\geq {T_i}+ \sum_{\ell \in \Psi_i} |\sbrackets{\vecth_i^{\text{eff}}}^H \vect{\omega}_\ell^c|^2 + \sum_{m \in \Omega_{i,k}} |\sbrackets{\vecth_i^{\text{eff}}}^H \vect{\omega}_m^c|^2-
      \nonumber \\
      &\frac{2\Re\cbrackets{ \sbrackets{\tilde{\vect{\omega}}_k^c}^H \vecth_i^{\text{eff}} \sbrackets{\vecth_i^{\text{eff}}}^H \vect{\omega}_k^c}}
      {\overset{}{\tilde{t}_k^c}}+ \frac{|\sbrackets{\vecth_i^{\text{eff}}}^H \overset{\phantom{I}}{\tilde{\vect{\omega}}_k^c}|^2}
      {\sbrackets{\tilde{t}_k^c}^2} \, t_k^c, \nonumber \\ &\hspace{6cm}\forall k \in \mathcal{K} ,\, \forall i \in \mathcal{M}_k .   \nonumber \\
     \label{eq:approxCon2}
\end{align}
Thus, we have the following convex optimization problem
 \begin{align}\label{eq:cmpW}
   & \underset{ \{\vect{w}_k,\vect{t}_k,\mat{R}_k\}_{k=1}^K}{\text{min}}\quad  \sum_{k\in \mathcal{K}} P(\vect{\omega}) \tag{P2.2}\\
   & \text{subject to} \qquad \nonumber\\
   &(\ref{eq:p2.1_first})-(\ref{eq:tkc}),\,(\ref{eq:approxCon1}),\,(\ref{eq:approxCon2}).\nonumber
\end{align}
Problem (\ref{eq:cmpW}) is convex and can be solved with standard optimization tools. Moreover, let $\mat{\Lambda} = \hbrackets{\vect{w}^T ,\,\vect{t}^T}^T$ be a vector stacking the optimization variables of (\ref{eq:cmpW}), $\widehat{\mat{\Lambda}}_z = \hbrackets{ \widehat{\vect{w}}_z^T ,\, \widehat{\vect{t}}_z^{\,\,T}\,}^T $ be the variables that are the optimal solution of problem (\ref{eq:cmpW}) computed at iteration $z$ and $\tilde{\mat{\Lambda}} = \hbrackets{ \tilde{\vect{w}}^T ,\, \tilde{\vect{t}}\phantom{.}^T\,}^T $ be the point, around which the approximations are computed. First, vector $\tilde{\mat{\Lambda}}$ is initialized by finding feasible \ac{MRC} beamformers $\tilde{\vect{w}}$ for the users. Using the initialization, can be solved to obtain the vector $\widehat{\mat{\Lambda}}_z$. If the current solution is not stationary,$\underset{\phantom{|}}{ }$ it is used for computing $\overset{\phantom{.}}{\tilde{\mat{\Lambda}}}$ for the next iteration, i.e. $\tilde{\mat{\Lambda}}_{z+1} = \tilde{\mat{\Lambda}}_z + \varrho_z \sbrackets{\widehat{\mat{\Lambda}}_z - \tilde{\mat{\Lambda}}_z},\,\text{for some }\varrho_z \in (0,1]$

\subsection{Phase Shift Design}
Given $\mat{\Lambda} = \hbrackets{\vect{w}^T ,\,\vect{t}^T}^T$, which is assumed to be fixed for the duration of optimizing the phase shift vector $\vectv$, Problem (\ref{eq:genProb}) can be expressed as a \ac{QCQP} problem {\cite{IRSImprovement1}. To this end, problem (\ref{eq:genProb}) can be reformulated into
\begin{align}\label{eq:genV-reform1}
   \text{find}\qquad &  \vect{v} \tag{P3.1}\\
   \text{subject to} \qquad
    &\gamma_k^p(\vectv) \geq t_k^p, && \forall k \in \mathcal{K} \label{eq:reform1_v-p}\\
    &\gamma_{i,k}^c(\vectv)\geq t_k^c,  && \forall k \in \mathcal{K} ,\, \forall i \in \mathcal{M}_k \label{eq:reform1_v-c}\\
    & | v_r | = 1,  && \forall r \in \{ 1,..., R\},
\end{align}
where the constraints (\ref{eq:reform1_v-p}) and (\ref{eq:reform1_v-c}) can be rewritten as
\begin{align}\label{eq:genV-reform2-p}
        &\hspace{-0.25cm}|\left( \vect{h}_k + \mat{H}_k \vect{v} \right)^H \vect{\omega}_k^p|^2 \geq t_k^p \left(
        \sum_{j\in\mathcal{K}\text{\textbackslash}\cbrackets{k}} \hspace{-0.175cm}|\left( \vect{h}_k + \mat{H}_k \vect{v} \right)^H \vect{\omega}_j^p|^2+ \right.\nonumber \\
        & \left.  \sum_{\ell \in \Psi_k} |\left( \vect{h}_k + \mat{H}_k \vect{v} \right)^H \vect{\omega}_\ell^c|^2 + \sigma^2\right), \hspace{1cm} \forall k \in \mathcal{K}
        \end{align}
        \begin{align}
         &\hspace{-0.2cm}|\left( \vect{h}_i + \mat{H}_i \vect{v} \right)^H \vect{\omega}_k^c|^2 \geq t_k^c\sbrackets{
         T_i + \sum_{\ell \in \Psi_i} |\left( \vect{h}_i + \mat{H}_i \vect{v} \right)^H \vect{\omega}_\ell^c|^2+ \right. \nonumber \\
         & \hspace{0.25cm} \left. \hspace{-0.2cm} \sum_{m \in \Omega_{i,k}} \hspace{-0.25cm} |\left( \vect{h}_i + \mat{H}_i \vect{v} \right)^H \vect{\omega}_m^c|^2},\forall k \in \mathcal{K} ,\, \forall i \in \mathcal{M}_k.\label{eq:genV-reform2-c}
\end{align}
This facilitates denoting
\begin{align}
   b_{k,i}^o &= \vect{h}_k^H \vectw_i^o, \text{ and }
   \vecta_{k,i}^o = \matH_k^H \vectw_i^o, \\
   \mat{M}_{k,j}^o &=
         \begin{bmatrix}
            \vect{a}_{k,j}^o ({\vect{a}_{k,j}^o})^H & \vect{a}_{k,j}^o (b_{k,j}^o)^H  \\
            {b_{k,j}^o} ({\vect{a}_{k,j}^o})^H  & 0  \\
         \end{bmatrix} \\
        \tilde{\vect{v}} \,&= \,[\vect{v}^T, s]^T,\label{eq:auxVar}
\end{align}
where $s$ is an auxiliary variable. Due to the auxiliary variable $s$, it holds that if a feasible solution $\vectvt^*$ is found, a solution $\vect{v}^*$ can be recovered by
$\vectv^*=[\vectvt^*/\tilde{v}^*_{R+1}]_{(1:R)}$, where $[\vect{x}]_{(1:R)}$ denotes the first $R$ elements of vector $\vect{x}$ and $x_r$ denotes the $r$-th element of vector $\vect{x}$ \cite{IRSImprovement1}.

To deal with the non-convex constraints, the matrix lifting technique can be applied to obtain a convex formulation \cite{federatedLearning}. Specifically, by defining $\matV = \vectvt\vectvt^H$, the phase shift vector $\vectvt$ is lifted into a \ac{PSD} matrix $\mat{V}$.

By expressing $\Tr{\matM_{k,i}^o \matV}\, \hat{=}\, \vectvt^H\matM_{k,i}^o\vectvt$ and introducing the slack variables $\zeta_k^p$ and $\zeta_k^c$ problem (\ref{eq:genV-reform1}) can be expressed as $\underset{\phantom{|}}{ }$
\begin{align}\label{eq:genV-alpha}
&\underset{\matV,\{\zeta_k^p,\zeta_{k}^c\}_{k=1}^K}{\text{maximize}}\qquad \sum_{k \in \mathcal{K}} \eta_k^p\zeta_k^p + \eta_k^c\zeta_{k}^c  \tag{P3.2}\\
   &\text{subject to} \qquad
    \sqr{|b_{k,k}^p|} +  \Tr{\matM_{k,k}^p\matV} \geq  t_k^p \nonumber \\& \Big(
        \sum_{j\in\mathcal{K}\text{\textbackslash}\cbrackets{k}} \sqr{|b_{k,j}^p|} +  \Tr{\matM_{k,j}^p\matV}
          \sum_{\ell \in \Psi_k} \sqr{|b_{k,\ell}^c|} +  \nonumber \\ & \Tr{\matM_{k,\ell}^c \matV}  + \sigma^2 \Big) + \zeta_k^p,  \hspace{0.5cm} \forall k \in \mathcal{K}  \label{eq:SINR_p-reform5}\\
    &\sqr{|b_{i,k}^c|}+  \Tr{\matM_{i,k}^c \matV} \geq t_k^c \Big(
         \sum_{j \in \mathcal{K}}^{\phantom{I}} \sqr{|b_{i,j}^p|} +  \Tr{\matM_{i,j}^p\matV} \nonumber \\ & + \sigma^2   + \sum_{\ell \in \Psi_i} \sqr{|b_{i,\ell}^c|} + \Tr{\matM_{i,\ell}^c \matV} + \sum_{m \in \Omega_{i,k}} \sqr{|b_{i,m}^c|} + \nonumber \\ &  \Tr{\matM_{i,m}^c \matV} \Big) + \zeta_{k}^c,  \hspace{0.35cm} \forall k \in \mathcal{K} ,\, \forall i \in \mathcal{M}_k  \label{eq:SINR_c-reform5}\\
    & V_{r,r} = 1, \hspace{3.25cm}\forall r \in \{ 1,..., R+1\}\\
    & \matV \succeq 0  \label{eq:genV-alpha_last}\\
    &  \rank{\matV} =1 \label{eq:genV-reform5_nonConvex},
\end{align}
where $\zeta^p_k$ and $\zeta_{k}^p$ can be interpreted as \textit{SINR residuals} of user $k$ in phase shift optimization \cite{IRSImprovement1} and \begin{align}
\eta_k^o = {\underset{i\in \mathcal{N}}{\sum}{\myNorm{\vectw_{i,k}^o}_2^2}}\big/{\underset{k\in\mathcal{K}}{\max}\cbrackets{\displaystyle\sum_{i \in \mathcal{N}}\myNorm{\vectw_{i,k}^p}_2^2 ,\displaystyle\sum_{i \in \mathcal{N}}\myNorm{\vectw_{i,k}^c}_2^2}}
\end{align}
are weighting factors that prioritize users, which require high transmission power.
To reformulate the rank-one constraint, the following proposition is introduced:
\begin{prop}\label{prop:1}
 \textit{For a \ac{PSD} matrix $\mat{X} \in \mathbb{C}^{N\times N}$ and }$\myNorm{\mat{X}}_* \geq 0$\textit{, it holds \cite{federatedLearning} that}
\end{prop}
\begin{align}\label{eq:eqR1}
   \text{rank}(\mat{X}) = 1 \Leftrightarrow \myNorm{\mat{X}}_* - \myNorm{\mat{X}}_2 = 0,
\end{align}
\textit{where $\myNorm{\mat{X}}_*$ denotes the nuclear norm and $\myNorm{\mat{X}}_2$ denotes the spectral norm of $\mat{X}$.}\\
\\
Similar to the lower bounds derived in (\ref{DC-approxp}) and (\ref{DC-approxc}), the first-order Taylor approximation of the spectral norm $\myNorm{\mat{V}}_2$ around point $\mat{V}^0$ can be derived as
\begin{align}\label{convCons1}
   \myNorm{\mat{V}}_2 \geq \myNorm{\mat{V}^0}_2 + \langle \partial_{\mat{V}^0} \myNorm{\mat{V}}_2 , \sbrackets{\mat{V}-\mat{V}^0}\rangle,
\end{align}
where $\partial_{\mat{V}^0}$ is the subgradient of $\myNorm{\mat{V}}_2$ with respect to $\matV$ at $\matV^0$ and where the inner product is defined as
$\langle \mat{X},\mat{Y} \rangle = \text{Re}\cbrackets{\Tr{\mat{X}^H\mat{Y}}}$,
as stated by Wirtinger's calculus \cite{wirtingers} in the complex domain. Therefore, a convex approximation of (\ref{eq:eqR1}) can be established, by replacing the spectral norm with the approximation given in (\ref{convCons1}). By adding the resulting expression as a penalty term to problem (\ref{eq:genV-alpha}) the following optimization problem can be obtained
\begin{align}\label{eq:genV-alpha_rank}
   &\underset{\matV,\{\zeta_k^p,\zeta_{k}^c\}_{k=1}^K}{\text{maximize}} \quad \rho \Big( \sum_{k \in \mathcal{K}}  \hspace{-0.05cm} \eta_k^p\zeta_k^p + \eta_k^c\zeta_{k}^c \Big) - \,  (1-\rho)  \tag{P3.3}\\ & \hspace{1cm} \Big(\hspace{-0.05cm}\myNorm{\mat{V}}_* - \myNorm{\mat{V}^0}_2 - \langle \partial_{\mat{V}^0} \myNorm{\mat{V}}_2 , \sbrackets{\mat{V}-\mat{V}^0}\rangle\hspace{-0.05cm} \Big) \nonumber \\
   &\text{subject to} \hspace{0.5cm}  (\ref{eq:SINR_p-reform5}) - (\ref{eq:genV-alpha_last}),
\end{align}
where $\rho$ is a trade-off factor, that regulates, the trade-off between a high quality solution and a rank-one solution. It is worth noting, that the subgradient $\partial_{\mat{V}^0}\myNorm{\mat{V}}_2$ can be efficiently computed by using the following proposition \cite{federatedLearning}.
\begin{prop}\label{prop:2}
\textit{For a given \ac{PSD} matrix $\mat{X}\in\mathbb{C}^{N\times N} $, the subgradient  $\partial_{\mat{X}}\myNorm{\mat{X}}_2$ can be computed as $\vect{e}_1\vect{e}_1^H$, where $\vect{e}_1 \in \mathbb{C}^N$ is the leading eigenvector of matrix $\mat{X}$.}
\end{prop}

Since the retrieved solution may not be rank-one, the Gaussian randomization technique is used to obtain a feasible solution to problem (\ref{eq:genV-alpha_rank}). To this end, the \ac{SVD} is applied to the solution $\matV^*$ of problem (\ref{eq:genV-alpha_rank}) as
$\matV^* = \mat{U} \mat{\Sigma} \mat{U}^H$,
where $\mat{U} \in \mathbb{C}^{(R+1)\times(R+1)}$ and $\mat{\Sigma} \in \mathbb{C}^{(R+1)\times(R+1)}$ are a unitary matrix and a diagonal matrix, respectively. Using these \ac{SVD} components, a rank-one candidate solution $\hat{\vectv}_g$ to problem (\ref{eq:genV-reform1}) can be generated \cite{GaussRandHelp} as
\begin{align}\label{eq:candidate}
   \hat{\vectv}_g = \mat{U}\mat{\Sigma}^{\frac{1}{2}} \vect{z}_g \in \mathbb{C}^{(R+1)\times 1},
\end{align}
where $\vect{z}_g \sim \mathcal{CN}(\zero_{R+1},\mat{I}_{R+1})$ denotes a random vector drawn independently from a circularly-symmetric complex Gaussian distribution. After generating $G$ randomized solutions, each randomization $\hat{\vectv}_g$ can be used to obtain a potential candidate solution $\overline{\vectv}_g$ for problem (\ref{eq:genV-reform1}), namely
\begin{align}\label{eq:finalVsol}
   \displaystyle \overline{\vectv}_g = \exp\left({\displaystyle j \arg{ \sbrackets{ \hbrackets{{\hat{\vectv}}/{\hat{v}_{R+1}}}_{(1:R)}}  }}\right).
\end{align} Each potential candidate solution $\overline{\vectv}$, that satisfies problem (\ref{eq:genV-reform1}), is a feasible candidate solution. Since there is no objective function in problem (\ref{eq:genV-reform1}), the highest achievable sum-rate is chosen as a criterion in order to find the best performing beamforming vector.

The proposed alternating optimization algorithm for solving problem (\ref{eq:genProb}) is outlined in Algorithm \ref{alg:final}, where problems (\ref{eq:genW}) and (\ref{eq:genV-alpha_rank}) are solved alternatively.
\begin{algorithm}
\footnotesize
\caption{Procedure to determine $\vectw^*$ and $\vectv^*$ of (\ref{eq:genProb})}\label{alg:final}
\begin{algorithmic}
\STATE \textbf{Input: } $\vectv_0$, $z\leftarrow0$,  threshold $\epsilon$, number of Gaussian randomizations $G$
\STATE Initialize: Find feasible $\widehat{\mat{\Lambda}}_0 = \hbrackets{({\vect{w}_0)}^T ,\,({\vect{t}_0)}^T  }^T$
\WHILE{the decrease of $P^{\text{Tr}}(\vect{\omega})$ is below $\epsilon$}  \IF{problem (\ref{eq:genV-alpha_rank}) is feasible}
   \IF{at least one feasible candidate $\overline{\vectv}_g$ can be determined}
      \STATE Set $\tilde{\mat{\Lambda}} \leftarrow \widehat{\mat{\Lambda}}_z$
      \STATE Obtain $\widehat{\mat{\Lambda}}_{z+1}$ by solving problem (\ref{eq:cmpW})
      \STATE Obtain $\vectv_{z+1}$ by solving problem (\ref{eq:genV-alpha_rank})
      \STATE $z \leftarrow z+1$
   \ENDIF
\ENDIF
\ENDWHILE
\STATE$\vectw^* \leftarrow \vectw_z$, $\vectv^* \leftarrow \vectv_z$
\STATE \textbf{Output: } Optimal $\vectw^*$ and optimal $\vectv^*$.
\end{algorithmic}
\end{algorithm}

\section{Numerical Simulations}\label{ch:numsim}
The simulation setup considers a \ac{C-RAN}, which consists of one \ac{CP}, that is connected to 4 \acp{BS} via fronthaul links. Each \ac{BS} is equipped with 4 antennas. The \acp{BS} operate in an area of $[-500,500] \times [-500,500] \text{ m}^2 $ and serve 6 single-antenna users. We also consider placing an \ac{IRS} in the center of this squared area, which consists of $R=15$$\underset{\phantom{|}}{ }$reflecting elements. The users and \acp{BS} are positioned uniformly and independently within the operation area. The channels between the \acp{BS}, users and \ac{IRS} follows the standard path-loss model consisting of three components: 1) path-loss as $\text{PL}_{x,y} = 148.1 + 37.6\log_{10}(d_{x,y})$, where $d_{x,y}$ is the distance between transmitter $x$ and receiver $y$ in km; 2) log-normal shadowing with 8dB standard deviation and 3) Rayleigh channel fading with zero-mean and unit-variance. The transmit bandwidth is assumed to be $B=10$ MHz, the noise power spectrum is set to -169 dBm/Hz and the minimum \ac{QoS} is set to $r_k^{\text{Min}}=4$ Mbps for each user. The sum-power weights $\alpha_k$ of the users are chosen in a uniform fashion, i.e. $\alpha_k \sim \mathcal{U}(1,2)$. It is assumed that the clusters are determined as $\mathcal{K}_n^p=\mathcal{K}_n^c=\mathcal{K},\, \forall n \in \mathcal{N}$. Furthermore, the number of Gaussian randomizations is set to $G=25$, the trade-off parameter is set to $\rho=0.9$ and $D=2$. 

Figure \ref{fig:normQoS} illustrates the normalized transmit power for different \ac{QoS} requirements of each user normalized to $r_k^{\text{Min}}$. The figure shows that the slopes of the lines, which represent the \ac{RS} schemes are smaller than the slopes of the lines of the respective \ac{TIN} scheme, which highlights the efficiency of the \ac{RS} scheme compared to \ac{TIN}. Deploying an \ac{IRS} lowers the required transmit power by around 5 dBm/Mbps if compared to the non-\ac{IRS} cases. Interestingly the performance gain of \ac{RS} compared with \ac{TIN} becomes larger when using an \ac{IRS}. Intuitively, with an \ac{IRS} the channel quality of all users significantly improves, however, this also leads to substantial interference increase. Hence, the RS gain becomes more pronounced in such a case.
\begin{figure}
\hspace{-0.65cm}\includegraphics[width = 1.15\columnwidth]{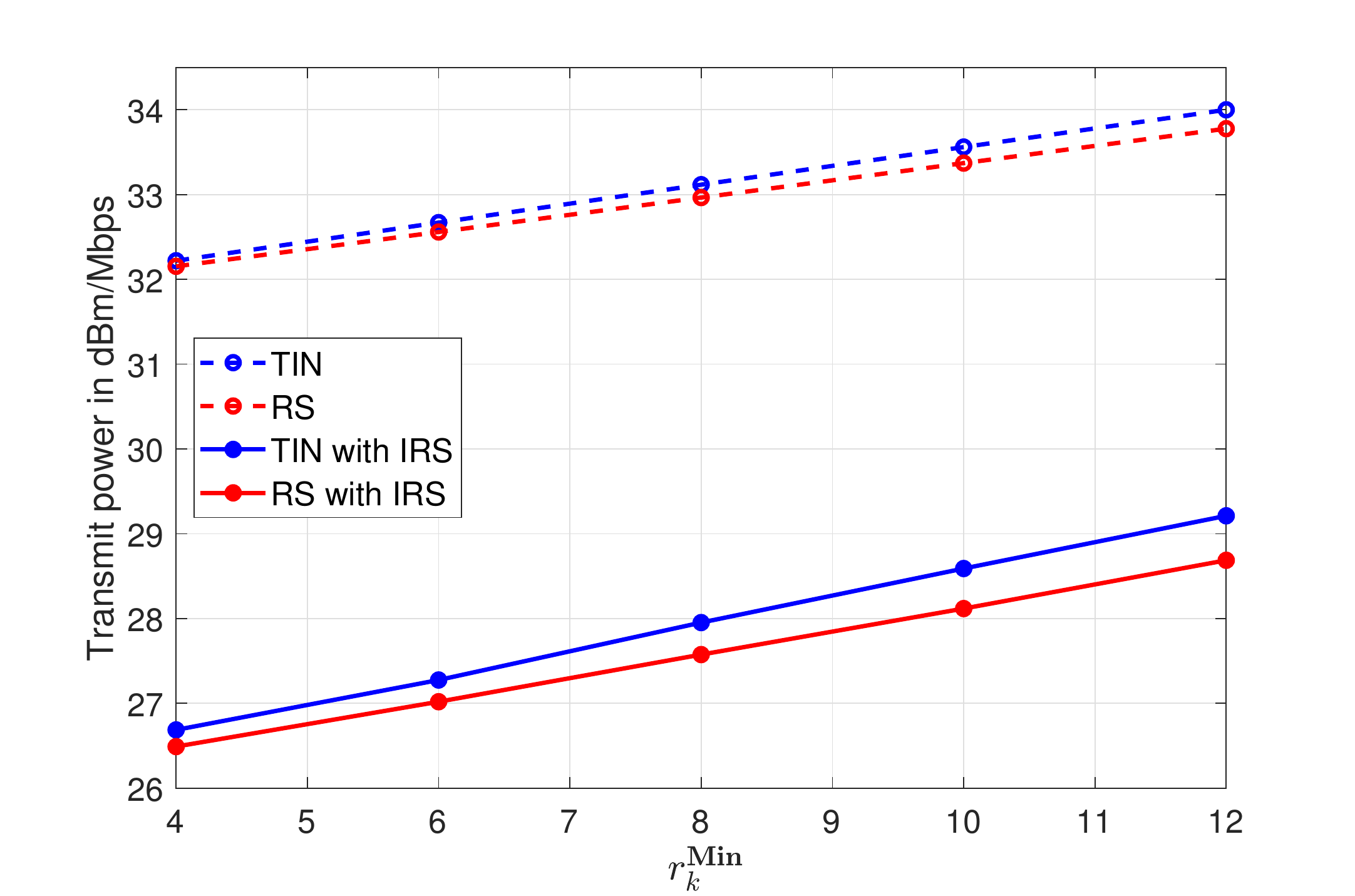}
\caption{The normalized minimum required transmit power to satisfy the \ac{QoS} requirements of the users.}
\label{fig:normQoS}
\end{figure}

\section{Conclusion}\label{ch:conc}
B5G networks are forecast to be dense networks that suffer from high interference but simultaneously require high data rates. With the goal to minimize the total transmit power consumption of the network, subject to per-user \ac{QoS} constraints, the \ac{IRS}-assisted \ac{C-RAN} is combined with the \ac{RS} scheme in order to mitigate the interference within the network. This also includes the interference amplified by the \ac{IRS}, which is caused by the optimization of the phase shifts towards the weaker users. To study the effect of this synergistic interaction between the \ac{RS} scheme and the \ac{IRS} on the reduction of the total required transmit power, an optimization problem is formulated. An alternating optimization approach is proposed to decouple the problem into two subproblems. As the resulting subproblems still remain non-convex, two frameworks are proposed, that obtain sub-optimal solutions to these subproblems. Results show that the combination of the \ac{IRS} with the \ac{RS} scheme can significantly decrease the required network power consumption as the \ac{RS} scheme is able to mitigate interference within the network, including the \ac{IRS}-assisted interference. In fact, the power savings, obtained by combining both techniques, exceed the sum of the power savings, that each technique can achieve individually. 
\balance

\footnotesize
\bibliographystyle{IEEEtran}
\bibliography{references}
\balance
%\section{Acronyms}
\begin{acronym}
\setlength{\itemsep}{0.1em}
\acro{AF}{amplify-and-forward}
\acro{AWGN}{additive white Gaussian noise}
\acro{B5G}{Beyond 5G}
\acro{BS}{base station}
\acro{C-RAN}{Cloud Radio Access Network}
\acro{CMD}{common message decoding}
\acro{CP}{central processor}
\acro{D2D}{device-to-device}
\acro{DC}{difference-of-convex}
\acro{IC}{interference channel}
\acro{i.i.d.}{independent and identically distributed}
\acro{IRS}{intelligent reflecting surface}
\acro{IoT}{Internet of Things}
\acro{LoS}{line-of-sight}
\acro{M2M}{Machine to Machine}
\acro{MIMO}{multiple-input and multiple-output}
\acro{MRC}{maximum ratio combining}
\acro{NLoS}{non-line-of-sight}
\acro{PSD}{positive semidefinite}
\acro{QCQP}{quadratically constrained quadratic programming}
\acro{QoS}{quality-of-service}
\acro{RF}{radio frequency}
\acro{RS-CMD}{rate splitting and common message decoding}
\acro{RS}{rate splitting}
\acro{SDP}{semidefinite programming}
\acro{SDR}{semidefinite relaxation}
\acro{SIC}{Successive Interference Cancellation}
\acro{SINR}{signal-to-interference-plus-noise ratio}
\acro{SOCP}{second-order cone program}
\acro{SVD}{singular value decomposition }
\acro{TIN}{treating interference as noise}
\acro{UHDV}{Ultra High Definition Video}

%\acro{M2M}{Machine to Machine}
%\acro{B5G}{Beyond 5G}
%\acro{CP}{Central Processor}
%\acro{IRS}{Intelligent Reflecting Surface}
%\acro{IoT}{Internet of Things}
%\acro{BS}{base station}
%\acro{C-RAN}{Cloud Radio Access Network}
%\acro{TIN}{treating interference as noise}
%\acro{RS}{rate splitting}
%\acro{CMD}{common message decoding}
%\acro{RS-CMD}{rate splitting and common message decoding}
%\acro{UHDV}{Ultra High Definition Video}
%\acro{LoS}{line-of-sight}
%\acro{NLoS}{non-line-of-sight}
%\acro{AF}{amplify-and-forward}
%\acro{RF}{radio frequency}
%\acro{QoS}{quality-of-service}
%\acro{QCQP}{quadratically constrained quadratic programming}
%\acro{DC}{difference-of-convex}
%\acro{IC}{interference channel}
%\acro{SIC}{Successive Interference Cancellation}
%\acro{AWGN}{Additive White Gaussian Noise}
%\acro{SINR}{signal-to-interference-plus-noise
%ratio}
%\acro{SINRs}{signal-to-interference-plus-noise
%ratios}
%\acro{MRC}{maximum ratio combining}
%\acro{D2D}{device-to-device}
%\acro{MIMO}{multiple-input and multiple-output}
%\acro{i.i.d.}{independent and identically distributed}
%\acro{SOCP}{second-order cone program}
%\acro{SDR}{semidefinite relaxation}
%\acro{SDP}{semidefinite programming}
%\acro{PSD}{positive semidefinite}
%\acro{SVD}{singular value decomposition}
\end{acronym}

\balance
\end{document}